\documentclass[useAMS,fleqn,usenatbib]{mnras}
\pdfoutput=1 
\usepackage{graphicx}
\usepackage{amsmath,amssymb,amstext}
\usepackage[T1]{fontenc}
\usepackage{newtxtext,newtxmath}
\usepackage[figure,figure*]{hypcap}
\input{hyperlink-year-only-natbib-patch}

\newcommand*{\kms}{\text{km\,s$^{-1}$}}
\newcommand*{\msun}{\text{M$_\odot$}}
\newcommand*{\perh}{\ensuremath{h^{-1}}}

\newcommand*{\mvir}{\ensuremath{M_\text{vir}}}
\newcommand*{\conc}{\ensuremath{c}}
\newcommand*{\spin}{\ensuremath{\lambda}}
\newcommand*{\ahalf}{\ensuremath{a_{1/2}}}
\newcommand*{\accrate}{\ensuremath{\Gamma_\text{peak}}}
\newcommand*{\nsub}{\ensuremath{N_\text{sub}}}
\newcommand*{\rmem}{\ensuremath{R_\text{mem}}}

\newcommand*{\vmax}{\ensuremath{V_\text{max}}}
\newcommand*{\mpeak}{\ensuremath{M_\text{peak}}}
\newcommand*{\vpeak}{\ensuremath{V_\text{peak}}}
\newcommand*{\zpeak}{\ensuremath{z_\text{peak}}}
\newcommand*{\spinb}{\ensuremath{\lambda_\text{Bullock}}}
\newcommand*{\almm}{\ensuremath{a_\text{LMM}}}
\newcommand*{\accratei}{\ensuremath{\Gamma_\text{inst}}}
\newcommand*{\subvpeak}{\ensuremath{V_\text{peak}^\text{(1st sub)}}}
\newcommand*{\rmemw}{\ensuremath{R_\text{mem}^\text{(weighted)}}}

\newcommand*{\mailto}[1]{\href{mailto:#1}{#1}}
\newcommand*{\http}[1]{\href{http://#1}{#1}}

\title[Beyond assembly bias]{Beyond assembly bias: exploring secondary halo biases for cluster-size haloes}
\author[Mao, Zentner \& Wechsler]{
Yao-Yuan~Mao$^{1,4}$\thanks{E-mail: \mailto{yymao.astro@gmail.com}, \mailto{yymao@pitt.edu}},
Andrew~R.~Zentner$^{1,4}$,
Risa~H.~Wechsler$^{2,3,4}$
\\
$^{1}$Department of Physics and Astronomy and the Pittsburgh Particle Physics, Astrophysics and Cosmology Center (PITT PACC),\\ 
\phantom{$^{1}$}University of Pittsburgh, Pittsburgh, PA 15260, USA\\
$^{2}$Kavli Institute for Particle Astrophysics and Cosmology and Department of Physics, Stanford University, Stanford, CA 94305, USA\\
$^{3}$SLAC National Accelerator Laboratory, Menlo Park, CA 94025, USA\\
$^{4}$Kavli Institute for Theoretical Physics, University of California, Santa Barbara, CA 93106, USA}
\pubyear{2018}

\begin{document}

\label{firstpage}
\pagerange{\pageref{firstpage}--\pageref{lastpage}}
\maketitle

\begin{abstract}
Secondary halo bias, commonly known as `assembly bias,' is the dependence of halo clustering on a halo property other than mass. This prediction of the $\Lambda$--Cold Dark Matter cosmology is essential to modelling the galaxy distribution to high precision and interpreting clustering measurements. As the name suggests, different manifestations of secondary halo bias have been thought to originate from halo assembly histories. We show conclusively that this is incorrect for cluster-size haloes. We present an up-to-date summary of secondary halo biases of high-mass haloes due to various halo properties including concentration, spin, several proxies of assembly history, and subhalo properties. While concentration, spin, and the abundance and radial distribution of subhaloes exhibit significant secondary biases, properties that directly quantify halo assembly history do not. In fact, the entire assembly histories of haloes in pairs are nearly identical to those of isolated haloes. In general, a global correlation between two halo properties does not predict whether or not these two properties exhibit similar secondary biases. For example, assembly history and concentration (or subhalo abundance) are correlated for both paired and isolated haloes, but follow slightly different conditional distributions in these two cases. This results in a secondary halo bias due to concentration (or subhalo abundance), despite the lack of assembly bias in the strict sense for cluster-size haloes. Due to this complexity, caution must be exercised in using any one halo property as a proxy to study the secondary bias due to another property.
\end{abstract}


\section{Introduction}
\label{sec:intro}

Hierarchical structure formation is one of the most profound predictions of the standard model of cosmology --- the Lambda--Cold Dark Matter ($\Lambda$CDM) cosmology \citep[e.g.,][]{10.1093/mnras/183.3.341,10.1086/183911,10.1038/311517a0,2010gfe..book.....M}. In a $\Lambda$CDM universe, dark matter is attracted to local density peaks of initial fluctuations, laid down during inflation, forming dark matter haloes. These dark matter haloes grow and merge with one another, and are believed to be the nests within which visible galaxies develop. The spatial distribution of dark matter haloes is hence a powerful prediction of the $\Lambda$CDM model, and can be compared with observations of galaxy distribution to test our understanding of cosmology and galaxy formation physics. 

Despite the nonlinearity in the formation and merging of dark matter haloes, we have a good general picture of the distribution of dark matter haloes, thanks to extensive studies using both mathematical approximations and numerical simulations. In particular, the excursion set formalism provides an analytic description of the halo mass function, halo mass assembly histories, and halo clustering properties  (\citealt{Press1974,1991ApJ...379..440B}; for a review see \citealt{10.1142/S0218271807010511} and Chapter 7 of \citealt{2010gfe..book.....M}). In this context, the clustering properties of haloes are a function of halo mass alone. This is commonly referred to as ``halo bias'' and can be approximated analytically \citep[e.g.,][]{Kaiser84,Cole1989,Mo1996,Sheth2001} or calibrated against numerical simulations \citep[e.g.,][]{Tinker2010}. These analytic descriptions of halo bias are powerful tools within galaxy models that rely on halo mass as the link between haloes and the galaxies that they host, such as the Halo Occupation Distribution \citep[e.g.,][]{Peacock2000,Seljak2000,Berlind2002,Zheng2005,Zehavi2005} and the Conditional Luminosity Function \citep[e.g.,][]{Yang2003,vandenBosch2013}, to predict galaxy clustering efficiently. 

On the other hand, it has also been shown, within cosmological $N$-body simulations, that halo clustering also depends on halo properties other than mass, amongst which the most notable is halo assembly history  \citep{2001PhDT.........7W,Gao2005,Wechsler2006,2008MNRAS.389.1419L}. This is commonly referred to as ``halo assembly bias.'' Although halo assembly bias is a smaller effect than mass-dependent halo bias, it has drawn increasing attention from researchers who model the galaxy--halo connection \citep[e.g.,][]{1507.01948,Hearin2016,Zentner2016,1611.09787,Lehmann2017,2017MNRAS.469.1809R}, because it is very plausible that the galaxy assembly history is to some extent connected to the assembly history of its host halo. If this is the case, one would need to understand halo assembly bias to predict accurately galaxy clustering and to mitigate potential bias in any inference from clustering measurements \citep{Reddick2013,Zentner2014}. 

In the excursion set formalism, a characteristic mass $M^*$, below which most haloes have already collapsed and formed, is defined to satisfy $\sigma(M^*)=\delta_c D(a)$, where $\delta_c \simeq 1.686$ is the critical overdensity, $D(a)$ is the linear growth rate, and $\sigma(M)$ is the squared root of the mass variance with a top-hat filter of mass $M$. 
Studies have found that, below the characteristic mass $M^*$, haloes that form earlier are more strongly clustered \citep{2001PhDT.........7W,Gao2005,Wechsler2006,2008MNRAS.389.1419L}. However, above the characteristic mass $M^*$, the signal of halo assembly bias is less clean. \citet{Wechsler2006} found that, above the collapse mass, haloes with lower \emph{concentration} are more clustered, but detected no clear bias as a function of halo formation time, despite the correlation between concentration and formation time \citep{Wechsler2002}. Similarly, \citet{2008MNRAS.389.1419L} inspected several different definitions of halo formation time, but found little assembly bias for high-mass haloes upon splitting haloes by their formation times. Attempts have been made to explain the physical origin of halo assembly bias \citep{Sandvik2007,10.1142/S0218271807010511,Desjacques2007,Dalal2008,2009MNRAS.396.2249W}, and while they provided some heuristic insights, they do not explain why different proxies of halo assembly history (e.g., concentration and formation time) exhibit assembly biases of very different magnitudes for massive haloes.

On the observational front, recent efforts have been made to detect halo assembly bias with galaxy clusters \citep{Yang2006,Tinker2012,Miyatake2015,More2016,Dvornik2017}. However, results remain inconclusive as various systematic effects can masquerade as observed halo assembly bias \citep{Lin2016,2016arXiv161100366Z,1702.01682}. One challenge facing observations is that, because halo properties are not directly observable, one has to use some observational proxy to approximate halo assembly history. In the studies of \citet{Miyatake2015,More2016,Dvornik2017}, they use the average distance of member galaxies in the cluster as a proxy of halo formation time. The motivation for such a proxy is the well-documented correlation between halo concentration and halo formation time \citep{Wechsler2002} and the assumed correlation between halo concentration and member galaxy distances. Nevertheless, even if a bias signal due to the average distance of member galaxies had been robustly detected observationally, it would not be clear whether or not this bias is of the same physical origin as the halo assembly bias identified in cosmological $N$-body simulations. 

In addition to halo assembly bias, the clustering of haloes also depends on other halo properties, such as spin, shape, and substructure abundance \citep[e.g.,][]{2001PhDT.........7W,Wechsler2006,Bett2007,Gao2007,2007MNRAS.375..489H,Faltenbacher2010}. {More recently, there are also studies that explore the relationships amongst different kinds of halo assembly bias \citep[e.g.,][]{1612.04360,1708.08451}.}  In much of the literature, these other \emph{secondary halo biases} (i.e., dependence of halo clustering on halo properties other than mass) are also commonly referred to as ``assembly bias,'' regardless of whether or not the secondary halo properties have a direct connection to halo assembly history. This nomenclature could result in some confusion as to whether all the different kinds of secondary halo bias have the same physical origin. While halo assembly history is arguably the most important property other than halo mass that characterizes individual haloes, when it comes to the clustering properties, it is still unclear if or how the different secondary halo biases connect with one another.

In this study, we join the exploration of halo secondary bias with two specific aims. First, we examine the dependence of halo clustering on a set of secondary halo properties for cluster-size haloes using a modern, large-volume cosmological simulation to validate previous results, some of which may have suffered from limited volume or resolution. Secondly, we inspect the correlations between different halo properties and how they connect to the secondary bias, to better understand the relation between assembly bias and all different kinds of secondary biases. To this end, we present a novel way to characterize secondary bias, which helps us to gain insight into these questions. 

We focus our current study on cluster-size haloes for three reasons. First, in the cluster-mass regime, the behaviour of assembly bias due to different proxies, such as halo concentration and formation time, is poorly understood. Secondly, at the high-mass end of the halo mass spectrum, halo assembly bias is thought not to be due to nonlinear interactions between neighbouring haloes, but reflective of the initial conditions for structure formation \citep{10.1142/S0218271807010511,Dalal2008}. Therefore, it seems plausible that assembly bias is simpler at high mass.
Third, there have been, and likely will soon be more, observational attempts to directly measure halo assembly bias with galaxy clusters. Hence, this is a timely and crucial study to facilitate the interpretation of current measurements and preparation for forthcoming observations.  

This paper is structured as follows. In \autoref{sec:methods}, we introduce the simulation used in this study, define the halo properties, and explain how we remove mass dependence in the bias. We present our main results in \autoref{sec:results}, where we show and compare the secondary halo biases due to different halo properties, with \autoref{fig:bias} summarizing this result, and also present a novel way to characterize secondary bias. In \autoref{sec:discussion}, we turn to explore the interplay between correlation and bias, demonstrate that the correlation between different halo properties does not determine the secondary bias they exhibit, and also discuss the implication for galaxy assembly bias. \autoref{fig:history} highlights the absence of assembly bias (in its strict definition) at this mass scale, despite the existence of other secondary halo biases. We conclude in \autoref{sec:conclusion}. We also include a summary of the correlations amongst all secondary halo properties used in this study and their dependence on the environment in Appendix~\ref{sec:appendix}.

\begin{figure*}
\centering\includegraphics[width=\textwidth]{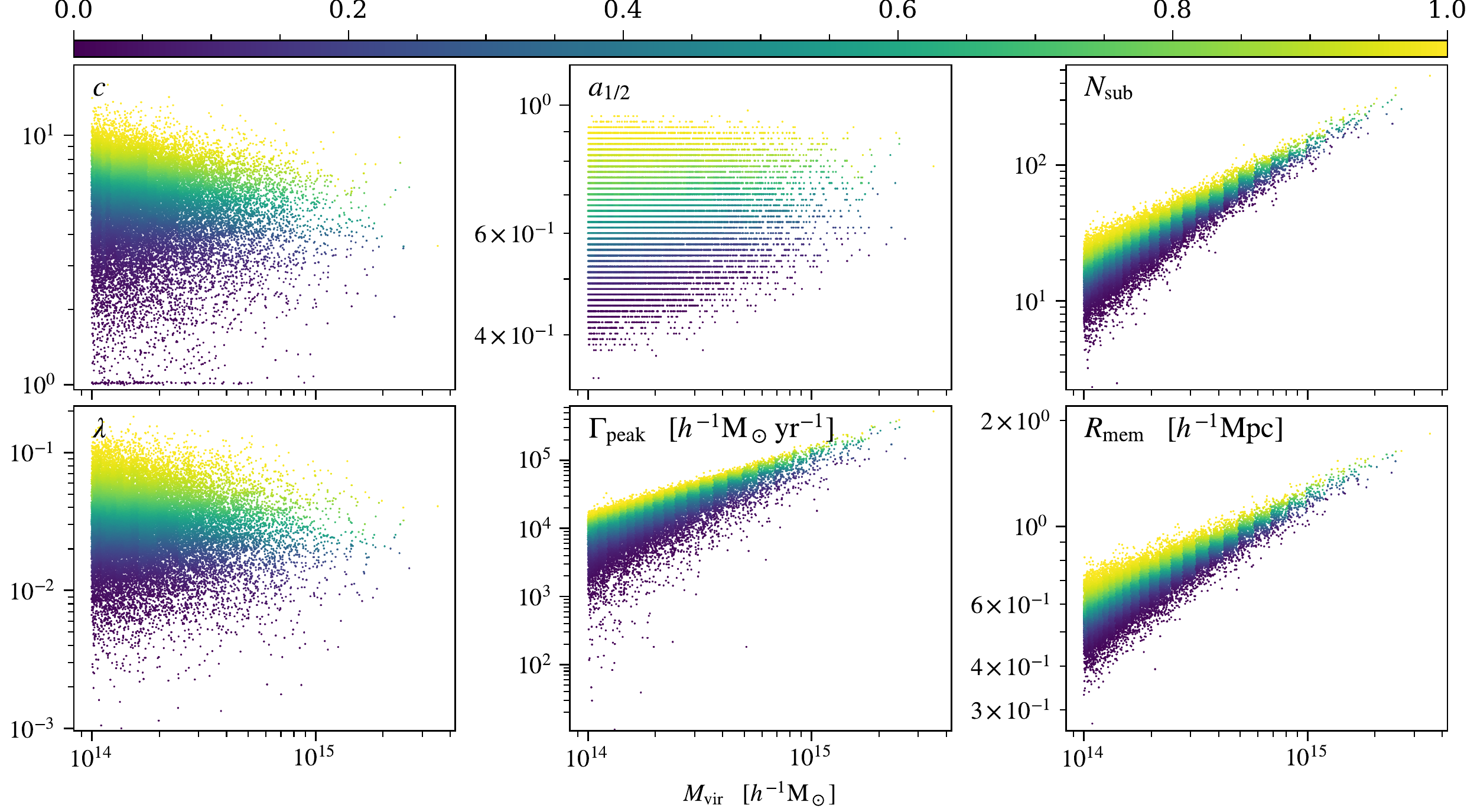}
\caption{\label{fig:marks}%
Each scatter plot shows the relation between halo mass and one of the secondary halo properties: concentration parameter \conc, spin parameter \spin, half-mass scale \ahalf, accretion rate before peak mass \accrate, number of subhaloes \nsub, and averaged subhalo distance \rmem. All distinct haloes in the sample are plotted and coloured by their mark values, which are always between 0 and 1 and are assigned according to the rank of the secondary property within each halo mass bin as explained in \autoref{sec:marks}.
}
\end{figure*}

\section{Methods}
\label{sec:methods}

\subsection{Simulations}
\label{sec:sims}

In this study we use the MultiDark Planck 2 (MDPL2) simulation \citep{Klypin2016}. MDPL2 is a cosmological gravity-only $N$-body simulation, run with the \textsc{L-Gadget2} code. It has a periodic volume of 1\,\perh\,Gpc$^3$, with 3840$^3$ particles. The mass resolution of each particle is $1.51\times10^9$\,\perh\,\msun, and the physical force resolution ranges from 5 to 13\,\perh\,kpc (smaller at lower redshift). 
MDPL2 adopts the Planck 2013 $\Lambda$CDM cosmology \citep{Planck2013cosmology}, and the actual values used in MDPL2 are: total matter density $\Omega_\text{m} = 0.307115$, dark energy density $\Omega_\Lambda = 1 - \Omega_\text{m}$, baryon density $\Omega_\text{b} = 0.048206$, Hubble parameter $h = 0.6777$, scalar spectral index $n_s = 0.96$, and the amplitude of mass density fluctuation $\sigma_8 = 0.8228$.

The MDPL2 simulation has been analysed by the \textsc{Rockstar} halo finder and the \textsc{Consistent Trees} merger tree builder \citep{2013ApJ...762..109B,2013ApJ...763...18B}. 
The halo mass definition used here is the virial mass; in this cosmology the virial overdensity corresponds to approximately 100 times the critical density \citep{Bryan1998}.
A halo is called a ``subhalo'' if its centre is within the virial radius of any larger halo. Any halo that is not a subhalo is called a ``distinct halo.''

In this study, we use haloes from the present-day ($z=0$) halo catalogue. We select all distinct haloes with a present-day virial mass $\mvir \geq 10^{14}$\,\perh\,\msun{} as our full halo sample. There are 27,029 distinct haloes in this sample, corresponding to a number density of $2.7\times 10^{-5}\,h^3\,\text{Mpc}^{-3}$.

Although the MDPL2 halo catalogues contain information about halo mass assembly histories, we do not have direct access to the full assembly history of each individual halo. Hence we also use the DarkSky-Gpc simulation when the full assembly history is required (\autoref{fig:history}). DarkSky-Gpc ({\tt ds14\_b}) is part of the Dark Sky Simulations \citep{Skillman2014}, run with the \textsc{2HOT} code \citep{Warren2013}. It also has a periodic volume of 1\,\perh\,Gpc$^3$, and was run with 10240$^3$ particles and a mass resolution of $7.63\times10^7$\,\perh\,\msun. 
DarkSky-Gpc has also been analysed by the \textsc{Rockstar} halo finder; however, the full \textsc{Rockstar--Consistent Trees} halo catalogues and merger trees have only been constructed on a downsampled version, which has only $10240^3/32 \simeq 3225^3$ particles and an effective mass resolution of $2.44\times10^9$\,\perh\,\msun%
\footnote{See also \citet{2016PhDT........76M,Lehmann2017}. We note that these two studies mistakenly reported the DarkSky-Gpc particle mass to be twice its actual value.}, similar to the mass resolution of MDPL2. In this study we only use the downsampled version of DarkSky-Gpc. DarkSky-Gpc adopts a flat cosmology close to the Planck 2013 $\Lambda$CDM cosmology, with $h = 0.688$, $\Omega_\text{m} = 0.295$, $n_s = 0.968$, $\sigma_8 = 0.834$. We also use the virial overdensity as the halo mass definition for DarkSky-Gpc.

While we present most of our result (all except for \autoref{fig:history}) using the MDPL2 simulation because it is publicly available and has slightly better mass resolution, we have verified that the DarkSky-Gpc simulation produces very similar result, with no qualitative difference and little quantitative difference. Our result holds in both simulations.

\subsection{Secondary halo properties}
\label{sec:properties}

In this study we select six different halo properties other than halo mass and investigate the secondary halo bias due to these properties. The properties we choose are as follows:

\begin{enumerate}
\itemsep0.5\baselineskip

\item Concentration parameter (\conc), obtained by fitting the dark matter density profile to a Navarro--Frenk--White (NFW) profile \citep{Navarro1996,Navarro1997}.

\item Spin parameter (\spin), as defined in \citet{Peebles1969}.

\item Half-mass scale (\ahalf), defined as the scale factor at which a distinct halo reaches more than or equal to half of its present-day ($z=0$) mass on its main branch.

\item Accretion rate {of} peak mass (\accrate), the average halo mass accretion rate (in \perh\,\msun\,yr$^{-1}$) {of peak mass, i.e., \[\left[\mpeak(z=0) - \mpeak(z=0.5)\right]/\left[t(z=0) - t(z=0.5)\right].\]} Since our halo sample contains the most massive haloes in the simulation, the majority (68.7\%) have $\zpeak = 0$ {(the redshift when peak mass takes place)}. The median \zpeak{} for haloes whose present-day mass is not peak mass (i.e., $\zpeak > 0$) is 0.093. Hence, for our halo sample, \accrate is basically the average halo mass accretion rate for $0 < z < 0.5$ .

\item Number of subhaloes (\nsub). We count the number of subhaloes (identified by the `upid' value in the \textsc{Rockstar--Consistent Trees} catalogue) that have a peak maximal circular velocity (\vpeak) above 135\,\kms. The peak maximal circular velocity is defined as the largest value of the maximal circular velocity on the main branch of the subhalo in consideration. 

\item Average subhalo distance (\rmem), defined as the average three-dimensional distance between all subhaloes and the centre of the main halo. The subhalo definition is the same as the definition when we calculate the number of subhaloes for each distinct halo.

\end{enumerate}

For \conc, \spin, \ahalf, and \accrate, we use the values in the \textsc{Rockstar--Consistent Trees} catalogue directly.

\subsection{Definition of mass-normalized marks}
\label{sec:marks}

Formally, we say a secondary halo bias exists if two groups of haloes that have the same halo mass but different values of a secondary propriety cluster differently. However, practically, because it is difficult to measure spatial clustering within an infinitesimally thin mass bin due to the finite number of haloes even within our large-volume simulations, we need to remove any bias that may be induced by halo mass when we measure the secondary bias. 

To do so, for each secondary halo property in consideration, we assign a ``mass-normalized'' mark value to each halo to represent the secondary halo property. We first bin haloes by halo mass, and then in each mass bin, we assign the mark value based on the rank within that mass bin. Hence, within each mass bin and also overall, the mark values always span $[0,1]$ uniformly. The mark values are always dimensionless.

For secondary halo properties that are discrete, such as the number of subhaloes and the half-mass scale (which is discrete because of the time interval between the simulation snapshots that are saved), if a group of haloes in a particular mass bin share the same exact value, they are randomly assigned different mark values in a way that still preserve the overall ranks. For example, when assigning the mark for the number of subhaloes, we add a random number drawn from a continuous uniform distribution on $[0, 1)$ to the number of subhaloes before the ranking process. In this fashion, haloes with one subhalo would have different mark values, but their mark values will always be smaller than the mark values of haloes with two subhaloes. This procedure hence ensures that the mark is uniformly distributed and that it is not clustered at a certain value, yet it does not introduce noise in the overall ranks.

The sample of distinct haloes used in this study spans the mass range of $14 \leq \log\,\left[M/(\perh\,\msun)\right] < 15.55$. We split this mass range into 30 bins. The bin widths increase quadratically in log mass, such that the lowest mass bins do not contain the vast majority of haloes in the sample. \autoref{fig:marks} shows the relations between the halo mass and each of the secondary halo properties we considered here. The mark value we assigned to each halo is represented by the colour of each point. We can observe the binning effect on the mark values for properties that change more rapidly with halo mass (\accrate, \nsub, and \rmem). Nevertheless, we have tested and find our results are insensitive to the binning schemes. When we use different binning schemes, including uniformly spacing in log mass and also different number of bins, the result still holds.

\section{Results}
\label{sec:results}

\subsection{Secondary halo bias manifested in the bias function}
\label{sec:bias}

\begin{figure}
\centering\includegraphics[width=\columnwidth]{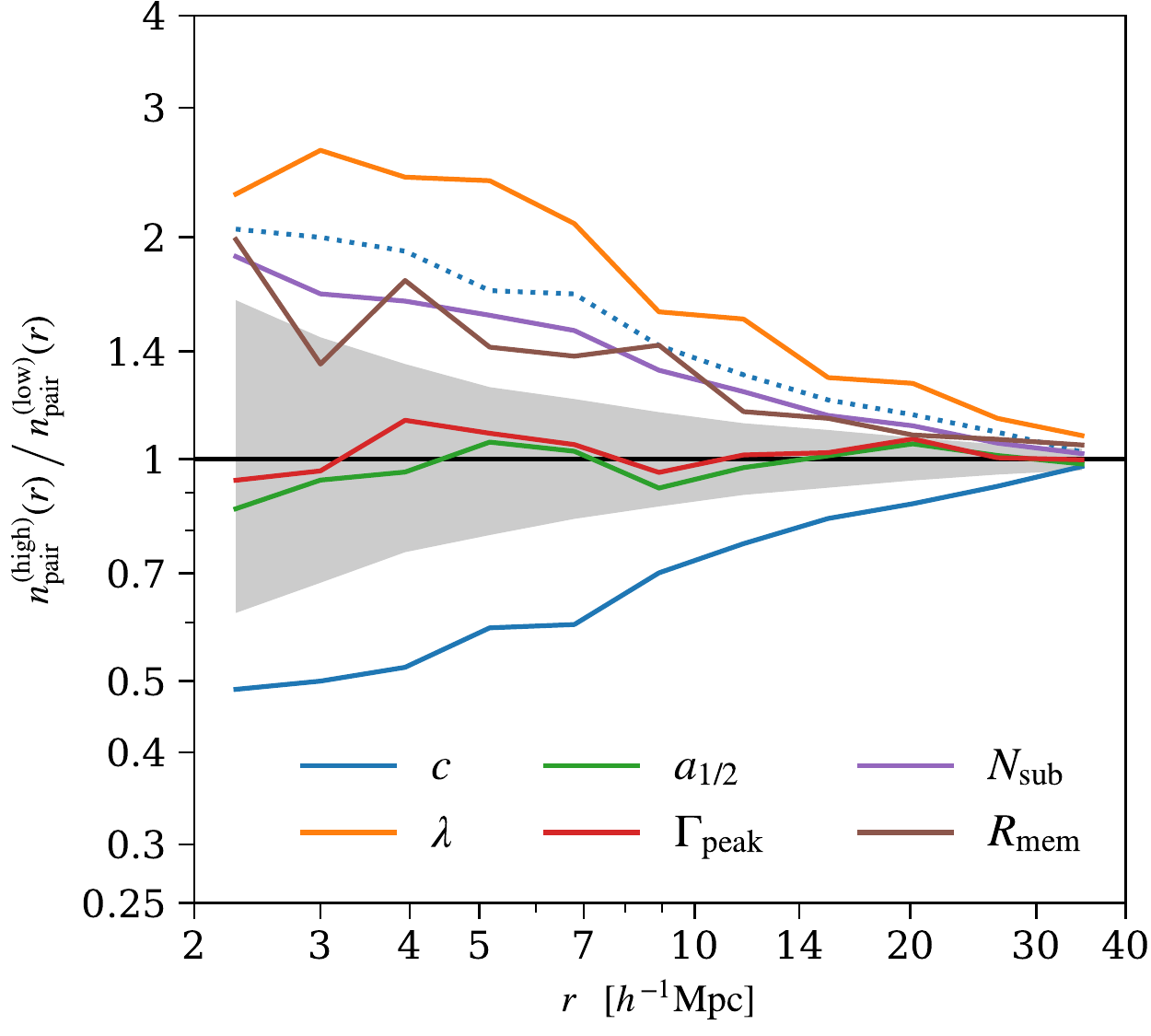}
\caption{\label{fig:bias}%
The ratio between the pair counting function ($n_\text{pairs}$) of only haloes with high marks and that of only haloes with low marks for each of the six secondary halo properties: (solid lines from top to bottom) \spin, \rmem, \nsub, \accrate, \ahalf, and \conc. A high (or low) mark value means a mark value above (or below) 0.5. The pair counting function is calculated by counting all pairs of haloes that have three-dimensional distances between 2 to 40\, \perh\,Mpc, and binned in distance ($r$), normalized by the square of number of haloes. For the concentration parameter (blue solid line), we also show the inverse ratio as a blue dashed line to guide an easy comparison with bias due to other secondary halo properties. For comparison, a thin horizontal black line shows a ratio of unity, and the grey band shows the 3$\sigma$ deviation for a randomly-assigned mark values. The deviation of the coloured lines from the horizontal black line shows a bias due to the secondary halo property.
}
\end{figure}

The most straightforward way to evaluate halo bias is to calculate the ratio between the pair counts of two samples of haloes. This is sometimes called the bias function. If the two samples have different numbers of total haloes, one needs to normalize the pair count first, usually by dividing out the expected number of pairs of a set of uniformly distributed random points. As for secondary halo bias, one can then calculate the bias function between two halo samples that differ in a secondary halo property, but not in halo mass. 

Here we present in \autoref{fig:bias}, for each of the six secondary halo properties listed in \autoref{sec:properties}, the bias function (or more precisely, the ratio between the pair counting functions) of two samples of haloes, split by the mark values. For each case, one sample has all haloes with mark values above 0.5, and the other has all haloes with mark values below 0.5. That is, we compare haloes in the upper half of the mark distribution with those in the lower half of the mark distribution at \emph{fixed} mass (cf., \autoref{fig:marks}). Note that because the mark values are assigned in mass bins, the split by mark already excludes any clustering dependence upon mass, and the bias we observe is the secondary bias.

The pair counting function is evaluated in bins of pair distance ($r$). Here we use 11 equally-spaced bins in logarithmic scale between 2 to 40\,\perh\,Mpc, and we have verified that our result is insensitive to the binning scheme.
We also estimate how much of the bias signal in the ratio of pair counting functions can come from noise. We uniformly at random assign mark values of $1/N, 2/N, \ldots, 1$ to all haloes in the sample, calculate the ratio of pair counting functions for high and low randomly assigned marks, and repeat this procedure until we obtain convergence of the 99.7\% (3$\sigma$) distribution of the bias signal; this is  shown by the grey band in \autoref{fig:bias}. Hence, deviation beyond this grey band indicates a secondary halo bias signal $>3\sigma$.

\begin{figure*}
\centering\includegraphics[width=\textwidth]{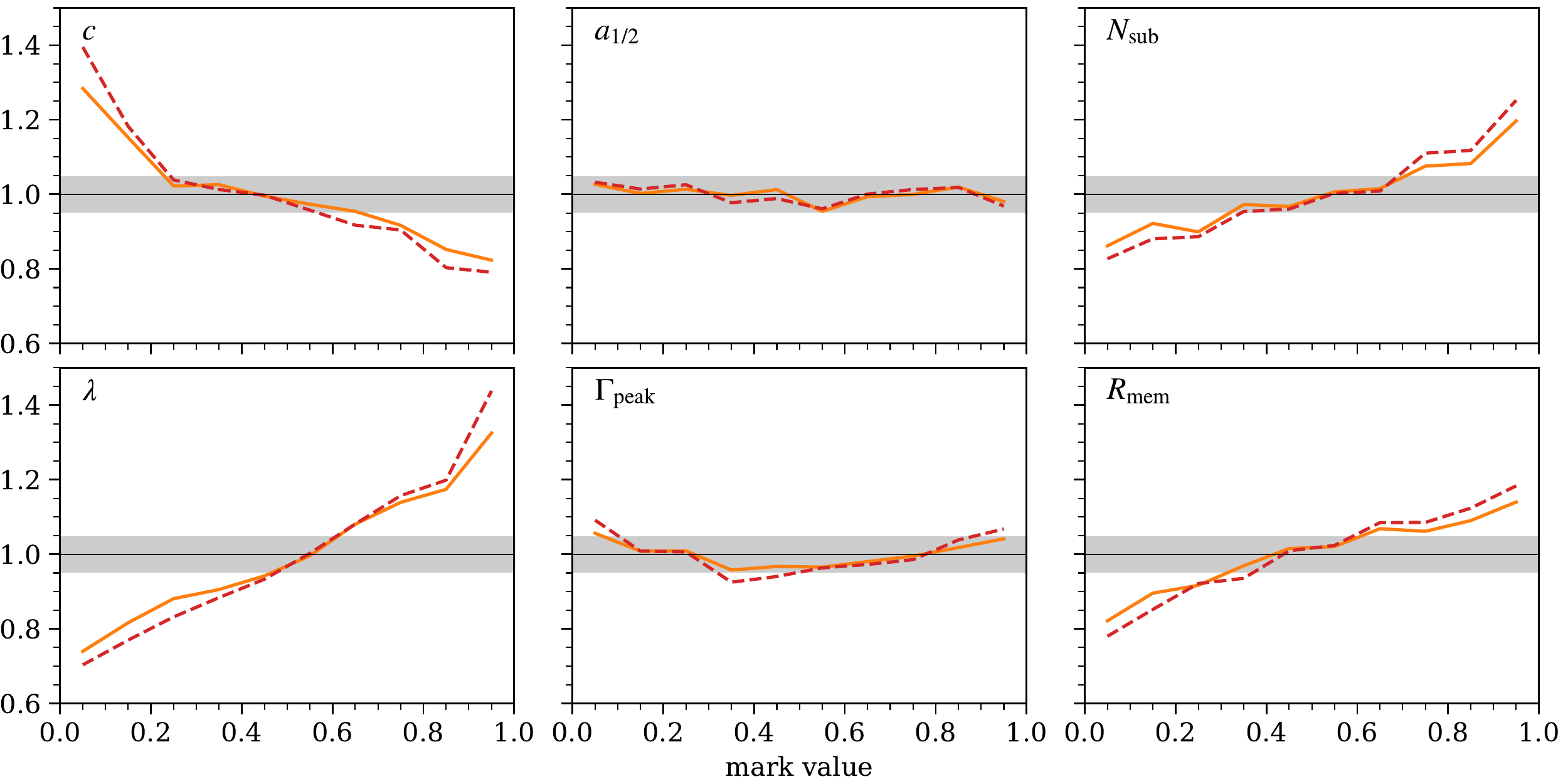}
\caption{\label{fig:markdist}%
The probability distributions of mark values of each of the six secondary halo properties, when only counting paired haloes. A ``paired halo'' is any distinct halo that is neighboured with at least one other distinct halo in the full halo sample within 10\,\perh\,Mpc. If a halo has in multiple neighbour haloes, its mark value is counted only once in the distribution shown by an orange solid line, and counted multiple times (as many as the number of its neighbours) in the distribution shown by a red dashed line. 
For comparison, a thin horizontal black line shows a uniform distribution, which, by construction, is the distribution of the mark values when including all, paired or not, haloes. The deviation of the coloured lines from the horizontal black line shows a bias due to the secondary halo property. The grey band shows a typical 3$\sigma$ deviation for a uniform random distribution of the same sample size.
}
\end{figure*}

\autoref{fig:bias} shows that, amongst the six secondary halo properties, the two that directly quantify some aspect of the mass accumulation histories of haloes, namely the half-mass scale $\ahalf$, and the accretion rate before peak mass $\accrate$, do not exhibit secondary bias; in both cases, the ratio between the the pair counts is consistent with random marks on all scales.

All other halo properties exhibit clear secondary bias. Haloes with higher concentrations are less clustered, while haloes with higher spins, more subhaloes, or a larger average subhalo distances are more clustered. In terms of the magnitude of the secondary bias, the spin parameter exhibits the strongest amongst these properties. The concentration parameter, number of subhaloes, and the average subhalo distance all produce a secondary bias of the roughly same magnitude. \autoref{fig:bias} also shows that, the secondary bias due to the properties listed decreases with scale. Nevertheless, between scales of 2 to 40\,\perh\,Mpc, this secondary bias is always statistically significant. Moreover, on all of the scales that we have considered, the secondary bias induced by any individual property never crosses the black horizontal line at unity, meaning that the secondary bias is always of the same sense. 

This result is broadly consistent with previous studies. In this mass regime, the inverted concentration bias has been shown in \citet{Wechsler2006}, the strong spin bias has been shown in \citet{Gao2007}, the lack of assembly bias when split by formation time has been shown in \citet{Gao2005,2008MNRAS.389.1419L}, and the bias due to subhalo distance has been shown in \citet{More2016}. Our study confirms that the secondary bias signals in previous studies can be reproduced with a larger-volume and higher-resolution simulation. {However, we caution the reader that we present the pair counting ratio in \autoref{fig:bias}, but this quantity 
does not directly translate into the ratio of linear biases of the subsamples. We will not quote linear biases in this work because the 
MDPL2 particle data are not publicly available.}

It is interesting to note that the term ``assembly bias'' has been widely use to refer to any secondary halo bias, despite the fact that {when splitting cluster-size haloes by their half-mass scales or recent accretion rates, the two sample have indistinguishable halo clustering properties}%
\footnote{We note that for low-mass haloes, which we do not address in this work, {there exists assembly bias even with this strict definition}.}.
The reason for this nomenclature is very likely due to the common understanding that halo assembly history is correlated with many halo properties, including concentration, spin, and so on. Nevertheless, the fact is that {some} direct measures of halo assembly history (e.g., $\ahalf$ and $\accrate$) result in no significant secondary bias at this mass regime, and this causes the ``assembly bias'' nomenclature to be potentially confusing and misleading. We will discuss this seemingly counter-intuitive result further in \autoref{sec:discussion}. 

\subsection{Secondary halo bias through the demography of paired haloes}
\label{sec:markdist}

\begin{figure*}
\centering\includegraphics[width=0.9\textwidth]{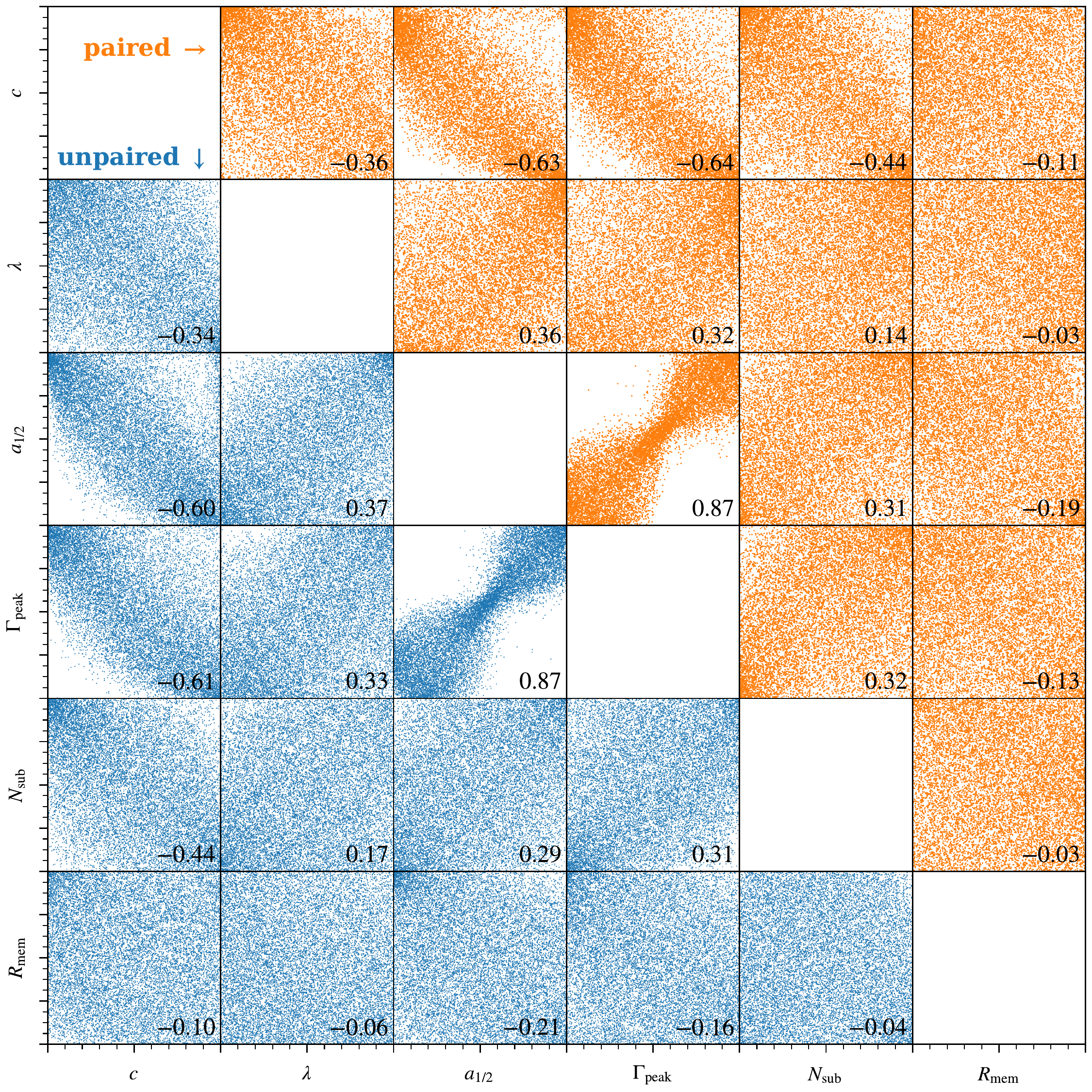}
\caption{\label{fig:scatter}%
A matrix of two-dimensional distributions of the mark values for each pair of the six secondary halo properties. In each cell, the $x$- and $y$-axes both go from 0 to 1 and show the mark value of the corresponding labels. The lower triangular cells of the matrix (with blue points) depict the distributions when only including haloes that are \emph{not} in pairs, while the upper triangular cells (with orange points) exhibit the distributions including only paired haloes. Paired haloes are those with at least one other distinct halo within 10\,\perh\,Mpc. The number in each cell is the value of the Spearman correlation coefficient. In the absence of correlations, the absolute value of Spearman correlation coefficient would be smaller than $0.03$ at the $99\%$ level  ($3\sigma$) for a sample the size of our halo sample. Diagonal cells are left blank as only a trivial perfect correlation would appear in those cells. For each cell in the upper triangular portion of the matrix, the marginal distribution of the points would match the orange solid lines in \autoref{fig:markdist}.
}
\end{figure*}

A different, yet arguably more intuitive, way to demonstrate the halo bias is to first split the halo sample by whether or not the halo contributes to the correlation (pair counting) function. Here we define a ``paired'' halo as any distinct host halo that is within 10\,\perh\,Mpc of another cluster-size distinct halo, and an ``unpaired'' halo is an isolated halo that has no other cluster-size distinct haloes within 10\,\perh\,Mpc.
The choice of 10\,\perh\,Mpc is, to some extent, arbitrary, but we have verified that our results are insensitive to the choice of this radius in the range of 5--20\,\perh\,Mpc. 
Once we divide the full halo sample into paired and unpaired subsets, we can inspect the difference between the demography of these two samples, and the difference is a manifestation of halo secondary biases.

We start with \autoref{fig:markdist}, showing the mark distribution of the six secondary halo properties for paired haloes only. If paired haloes form an unbiased subset of all haloes, the mark distribution of only paired haloes should be identical to the distribution of the full sample, which \emph{by construction} is uniformly distributed between $[0,1]$. 
Consequently, the deviations from the uniform distributions in \autoref{fig:markdist} highlight the secondary halo bias. Here we find that paired haloes are strongly biased towards lower concentration and higher spin, somewhat biased towards higher number of subhaloes and larger average subhalo distance, and nearly unbiased in half-mass scale and accretion rate before peak mass. All signals are consistent with the secondary bias signals in \autoref{fig:bias}. To verify that these signals are robust, we also show the mark distribution for all paired haloes with repeated counting, to account for the fact that some haloes contribute to multiple pairs and those haloes contribute more to the bias function. Nevertheless, in both counting schemes, the results are generally the same.

Note that this ``demography of paired haloes'' approach, demonstrated with \autoref{fig:markdist}, contains similar, but not identical, information to the bias function approach shown in \autoref{fig:bias}. For example, the cross-correlation between high- and low-mark haloes does not contribute to \autoref{fig:bias}, while any halo in pairs contributes to \autoref{fig:markdist}. Furthermore, \autoref{fig:markdist} helps us to understand the demography of paired haloes by allowing us to inspect the actual mark distribution for paired haloes. We can see that, for instance, the distributions of marks that exhibit secondary bias, are generally monotonic with respect to the mark values. This feature indicates that the secondary bias is not due to intermediate mark values but mostly due to high- or low-end tails in the mark distribution. In the next section we will further see the usefulness of this ``demography of paired haloes'' approach.

\section{Discussion}
\label{sec:discussion}

\subsection{Correlations between halo properties}
\label{sec:scatter}

The term ``halo assembly bias'' has been widely used to refer to any kind of secondary halo bias. While this is an issue of nomenclature, the extensive use of the term ``halo assembly bias'' has a significant drawback, as it implicitly suggests that all secondary halo bias may have originated from differences amongst the assembly histories of haloes. However, for haloes with the characteristic masses of galaxy clusters, we do not find significant secondary bias for \ahalf{} and \accrate{}, the two halo properties that we study which are directly related to halo assembly histories. This finding is in broad agreement with the detailed findings in the previous literature, though not always with the physical interpretation of them. In particular, despite the lack of assembly bias in its strict sense, there is still clear secondary halo bias for other properties that are correlated with halo assembly history, such as halo concentration. How can we understand better these counter-intuitive and seemingly contradictory results? 

We start by inspecting the correlations amongst the secondary halo properties. \autoref{fig:scatter} shows the scatter plot and the Spearman correlation coefficient between the mark values for each pair amongst the six secondary properties. \autoref{fig:scatter} shows these distributions for two subsets of haloes, namely paired haloes only (upper triangular cells in orange) and unpaired haloes only (lower triangular cells in blue). In order to interpret \autoref{fig:scatter}, it is useful to consider that in the absence of correlations, the Spearman correlation coefficient for a sample of the size of our halo subsets would be limited to an absolute value less than 0.03 at the 99.7\% ($3\sigma$) level. Therefore, for example, the correlation coefficient of $-0.06$ describing the correlation between $\lambda$ and $\rmem$ is a weak, but likely real, correlation. Correlation coefficients with a magnitude larger than this are very highly significant.

At first glance, most correlations are as expected based upon the previous literature. The half-mass scale \ahalf{} and the accretion rate before peak mass \accrate{} are strongly correlated, because the assembly histories of haloes are rather universal and in most cases can be well described by a one-parameter function \citep{Wechsler2002,Wu2013}. 
Halo concentration \conc{} is well correlated with both \ahalf{} and \accrate{}, consistent with the finding of \citet{Wechsler2002}. Halo concentration \conc{} is also fairly correlated with the number of subhaloes \nsub{}, consistent with the finding of \citet{Zentner2005,Mao2015}. The halo spin parameter \spin{} is correlated with \conc{}, \ahalf{}, and \accrate{}, also consistent with the finding of \cite{2007MNRAS.378...55M}. 
The average subhalo distance does not exhibit strong correlations with the other five halo properties, but correlates somewhat weakly with \ahalf{} and \accrate{}; this is consistent with the findings of \citep{More2016}.

A more unexpected feature of \autoref{fig:scatter} is that the correlations amongst these halo properties \emph{seem} to be nearly the same for paired and unpaired haloes, in terms of both the features in the scatter plots and the values of the Spearman correlation coefficients. This may be counter-intuitive as one might naively expect that if two properties are well correlated (such as concentration and half-mass scale), then both of them should result in secondary biases of roughly the same magnitude. However, this naive expectation is mathematically unfounded. As we demonstrate next, the distribution of the orange points (paired haloes) differs slightly from the distribution of the blue points (unpaired haloes). Because of this small difference, the correlation between two secondary properties, say concentration and half-mass scale, is far from a guarantee of that these two properties would result in similar secondary biases.

\subsection{Correlation does not imply secondary bias}
\label{sec:demo}

\begin{figure}
\centering\includegraphics[width=\columnwidth]{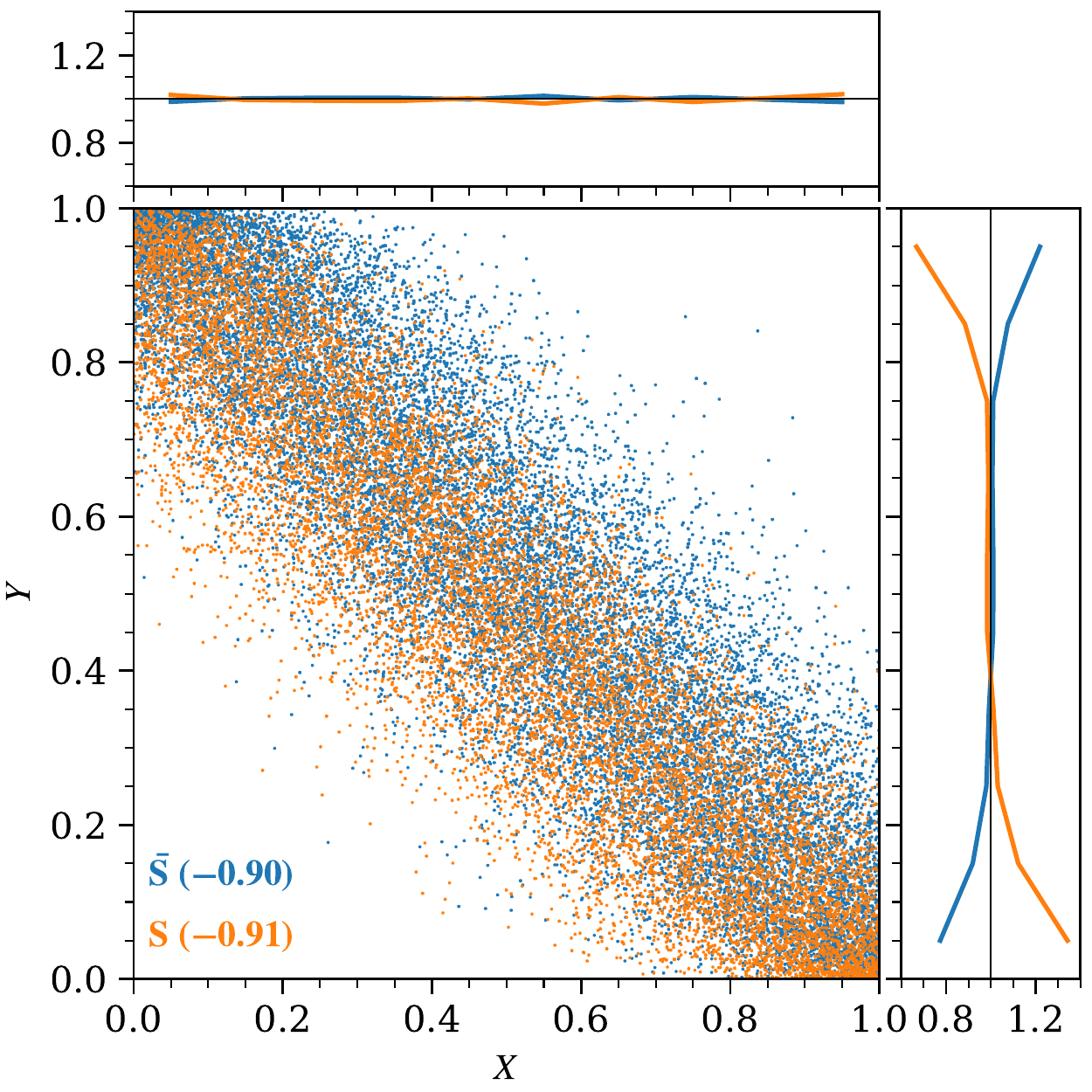}
\caption{\label{fig:demo}%
An illustration of two highly correlated variables that result in different biases when a subset is selected. The main panel shows the two-dimensional distribution (corresponding to \autoref{fig:scatter}) of two fictitious variables $X$ and $Y$, both uniformly distributed between $[0,1]$ for all points. The top and right-hand panels show the marginal probability distributions (corresponding to \autoref{fig:markdist}) of $X$ and $Y$, respectively. A subset of points (shown in orange, approximately 40\% of the total points) exhibit bias in $Y$ but not in $X$.  The numbers in parentheses are the Spearman correlation coefficient.
}
\end{figure}

\begin{figure*}
\centering\includegraphics[width=0.8\textwidth]{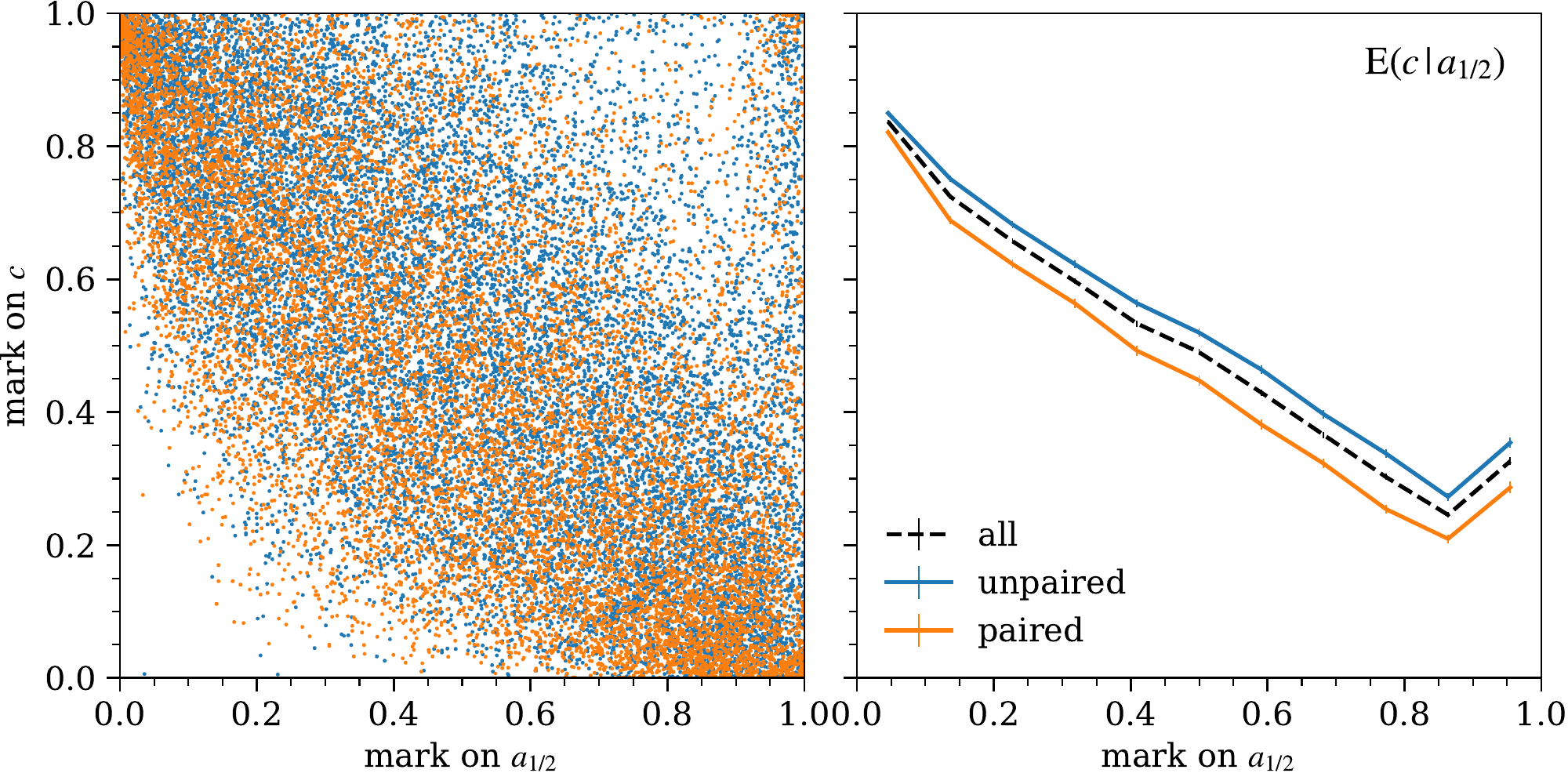}
\caption{\label{fig:c-ahalf}%
Left panel shows the two-dimensional distribution of the marks on the concentration parameter (\conc{}) and the half-mass scale (\ahalf{}), for unpaired (blue points) and paired (orange points) haloes. The distributions are the same as those in \autoref{fig:scatter}, but here we overlay them for a direct comparison. Right panel shows the mean value of the \conc{}-mark, conditioned on the \ahalf{}-mark; the black dashed, blue solid, and orange solid lines show the sample of all, paired, and unpaired haloes, respectively. 
The standard error of the mean is shown by error bars, but too small to be seen.
}
\end{figure*}

To demonstrate this important statement that two highly correlated halo properties can result in distinctly different secondary biases, consider a fictitious sample of 30,000 points that are described by two highly correlated variables $X$ and $Y$. The variables $X$ and $Y$ yield a Spearman correlation coefficient of $-0.90$, far more significant than the correlations amongst any of our halo secondary properties aside from the correlation of $\ahalf$ with $\accrate$. From amongst these 30,000 points, we select approximately 40\% of the total points and put them in a subset $S$. \autoref{fig:demo} shows the two-dimensional and marginal distributions of $X$ and $Y$, for both points in and out of the subset $S$. We can see that the subset $S$ exhibits bias in $Y$ but not in $X$, despite the very strong correlation between $X$ and $Y$. Of course, the selection of the points in $S$ is not random. The subset $S$ is constructed by preferably selecting points with lower $Y$ in bins of $X$, and hence it is by construction that this subset results in a different bias in $X$ and $Y$. Mathematically speaking, for the full sample we have $P(X)=P(Y)=1$ and $P(X,Y)=P(X|Y)=P(Y|X)$. However, for the subset $S$, we alter the conditional distribution $P_{S}(Y|X)$ so that it differs from $P_{S}(X|Y)$, and hence the marginal distributions $P_S(X)$ and $P_S(Y)$ differ.

The two-dimensional distribution in \autoref{fig:demo} is an analogy for \autoref{fig:scatter}, and the marginal distributions in \autoref{fig:demo} are analogies for \autoref{fig:markdist}. The subset of points $S$ is analogous to the subset of paired haloes. In fact, the marginal distributions for each cell in the upper triangular part of \autoref{fig:scatter} exactly corresponds to the mark distributions of paired haloes, shown by the orange solid lines in \autoref{fig:markdist}. Hence, even though by eye the two-dimensional distributions of two marks for paired and unpaired haloes look very similar, the small difference between them can result in markedly different secondary clustering biases.

To take an even closer look at the particular case of halo concentration and half-mass scale, we overlay the two-dimensional distributions of the mark values of halo concentration and half-mass scale for paired and unpaired haloes, and show them in the left-hand panel of \autoref{fig:c-ahalf}. Clearly, $\ahalf$ and $\conc$ are similarly correlated for both subsamples. Indeed, they have Spearman correlation coefficients that are consistent with each other given the sample size. However, we can already see the small difference between the two-dimensional distributions for paired and unpaired haloes by eye. To quantify this difference, in the right-hand panel of \autoref{fig:c-ahalf}, we show the mean value of the concentration mark, conditioned on half-mass scale mark, for the sample of all, paired, and unpaired haloes.
We find that, although concentration and half-mass scale are similarly correlated for paired and unpaired haloes, paired haloes have a slightly lower concentration at a given half-mass scale than unpaired haloes, and this small difference gives rise to their different behaviours in secondary bias.

At this point, it may be useful to summarize the phenomenology of \conc{}- and \ahalf{}-dependent halo clustering. Haloes exhibit secondary bias based on \conc{} because paired haloes constitute a subset of haloes with preferentially lower concentrations (in the mass range we consider). However, this subset of paired haloes also has the \emph{same} distribution of \ahalf{} as unpaired haloes (as shown in \autoref{fig:markdist}). Consequently, high-mass haloes exhibit \conc{}-dependent secondary bias, but do not exhibit $\ahalf$-dependent (or $\accrate$-dependent) secondary bias despite \ahalf{} and \conc{} being strongly correlated. Selecting haloes based upon whether or not they have a neighbour alters the conditional distributions $P(\conc|\ahalf)$ and $P(\ahalf|\conc)$, in a highly non-trivial manner. The underlying physical reasons for this shift in halo properties induced by proximity to neighbour haloes are not immediately apparent, but our work suggests that halo properties are related to environment in a manner that is considerably more complex than is commonly assumed.

Interestingly, the story of $\rmem$ is nearly the opposite of that for \ahalf{}. While \rmem{} exhibits a secondary bias similar to \conc, \spin, and \nsub{} (see \autoref{fig:markdist}), it is only weakly correlated with those three properties (see \autoref{fig:scatter}), for both paired and unpaired haloes. The lack of correlation of two variables by no means implies they cannot have similar or even the same marginal distribution, which is indeed what happens here. For example, the mark distributions of \rmem{} and \nsub{} for paired haloes are very similar, but these two properties are also the least correlated amongst the properties studied here. 

As a consequence, one should be very cautious when studying the secondary bias due to one halo property through the use of a different halo property as a proxy. Even when the two properties, namely the halo property of interest and the proxy, have been shown to be strongly correlated, it is not true that the two properties must exhibit similar secondary biases. Likewise, when two properties both exhibit similar secondary biases, they may still have little correlation with each other.

\subsection{Implication for galaxy assembly bias}
\label{sec:galaxy-bias}

The term ``assembly bias'' is also frequently used to refer to ``galaxy assembly bias,'' which does not have a single clear definition. In most contexts galaxy assembly bias means that the clustering properties of galaxies depend on some galaxy properties at a fixed host halo mass. With what we have learned here, we shall take a closer look at the idea of galaxy assembly bias. Consider a galaxy property $G$, which depends on a halo property $H$. If $G$ and $H$ are highly correlated and their conditional distributions $P(G|H)$ and $P(H|G)$ are not altered by the presence of nearby haloes (e.g., in the context of our examples, $P(G|H)$ and $P(H|G)$ stay the same for both paired and unpaired haloes), then the secondary halo bias due to $H$ induces a ``galaxy assembly bias'' due to $G$. However, if the conditional distributions $P(G|H)$ and $P(H|G)$ are altered for haloes in pairs, such as $(X,Y)$ and $(\ahalf,\conc)$ in our examples above, then the secondary halo bias due to $H$ does \emph{not} guarantee any bias signal due to $G$. Similarly, any bias signal due to $G$ cannot be used to infer the existence of an underlying secondary halo bias due to $H$. Likewise, seeing both biases due to $G$ and $H$ does not guarantee a correlation between the galaxy property $G$ and the halo property $H$.

In short, the correlation between two variables and the dependence of clustering properties on these two variables do not have a firm connection. This statement is particularly evident for the secondary halo biases for cluster-size haloes, but it is a general, mathematical statement. Hence, using ``assembly bias'' to refer to different kinds of secondary halo bias and even bias in galaxy clustering is potentially misleading. 

\subsection{Physical robustness of the secondary halo bias signals}
\label{sec:robust}

Despite some of the counter-intuitive and seemingly contradictory results that we have presented, the secondary halo biases we show in this work are, in fact, physically robust. One may suppose that either \ahalf{} or \accrate{} or both do not exhibit significant secondary halo clustering bias because they are measured with considerably more noise than the other halo properties that we explore or because these measures do not probe the particularly important epochs of halo formation. As we will show, this is not the case. Moreover, the existence of other secondary halo biases (due to concentration, spin, and subhalo properties) is also physically robust. 

To demonstrate the robustness of the secondary halo biases that we present above we proceed as follows. For each of the six secondary properties, we identify another, similar property that has approximately the same physical meaning but defined differently. We then investigate whether or not this similar halo property exhibits a similar secondary halo bias. This new set of six properties are as follows.

\begin{enumerate}
\itemsep0.5\baselineskip

\item Maximal circular velocity (\vmax). For a halo that follows a NFW profile perfectly, the maximal circular velocity is a simple function of halo mass and concentration. At fixed mass, higher \vmax{} implies higher halo concentration.

\item Spin parameter (\spinb), as defined in \citet{Bullock2001}, which has a different normalization compared to the Peebles spin parameter.  

\item Scale of last major merger(\almm), defined as the scale factor at which the halo experiences its last major merger on its main branch. A major merger is defined as a merging event of two haloes with a mass ratio that is larger than one-third. 

\item Present-day instantaneous accretion rate (\accratei), it is defined as the mass change rate between two adjacent snapshot outputs on the main branch. For MDPL2, this rate is calculated between $z=0$ and 0.0224, which corresponds to approximately 300\,Myr.

\item Peak maximal circular velocity of the largest subhalo (\subvpeak). Here largest subhalo means the subhalo that has the largest \vpeak{} values of all subhaloes in that distinct halo. At a fixed distinct halo mass, this value is correlated with the concentration of the host halo and with the number of subhaloes \citep{Mao2015}. 

\item Average subhalo distance weighted by subhalo mass (\rmemw), defined as the average three-dimensional distance between all subhaloes and the centre of the main halo, with the contribution of each subhalo weighted by the subhalo mass. The subhalo definition is the same as the definition used to calculate the number of subhaloes for each distinct halo.

\end{enumerate}

\begin{figure}
\centering\includegraphics[width=\columnwidth]{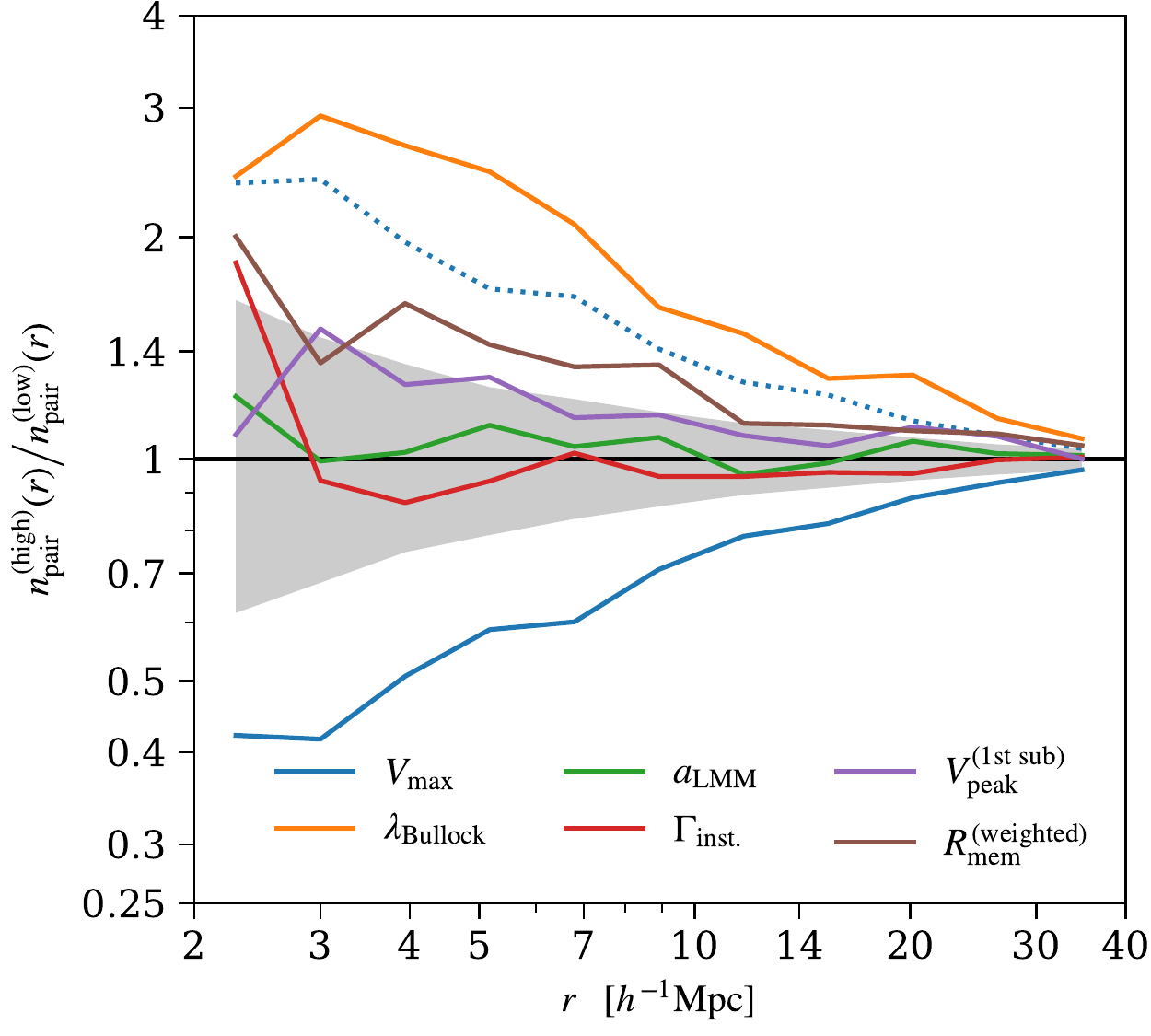}
\caption{\label{fig:bias-more}%
Same as \autoref{fig:bias} but for six different (but related) secondary halo properties: (solid lines from top to bottom) \spinb, \rmemw, \subvpeak, \almm, \accratei, and \vmax.
For \vmax (blue solid line), we also show the inverse ratio as a blue dashed line to guide an easy comparison with bias due to other secondary halo properties.
}
\end{figure}

\autoref{fig:bias-more} shows the secondary bias (as ratio of pair counting functions) for these six secondary properties, and the behaviours of the secondary biases are very similar to the six properties used in \autoref{fig:bias}. The spin parameter with \citet{Bullock2001} definition still exhibits the largest bias. The maximal circular velocity exhibits the second most significant secondary bias, similar to, but slightly stronger than, the secondary bias exhibited by the concentration parameter. The two assembly history-related properties (\almm{} and \accratei{}) again exhibit clustering that is consistent with random sampling and do not indicate secondary halo bias based upon halo mass assembly history. Lastly, the two subhalo-related properties (\subvpeak{} and \rmemw{}) show moderate secondary bias. 

As our discussion in \autoref{sec:demo} points out, correlation does not imply similar clustering bias. Hence the similar secondary bias signals we observe in \autoref{fig:bias} and \ref{fig:bias-more} can be interpreted to indicate that both properties in each pair affect the clustering properties. It also implies that the two conditional distributions of each pair of properties stay roughly the same for paired and unpaired haloes. In other words, the small change in definitions, in these particular cases, does not alter the joint distribution in a way that would result in different marginal distributions for paired or unpaired haloes (see \autoref{fig:conditional-mean} for a full comparison of the conditional distributions amongst all properties).

\begin{figure*}
\centering\includegraphics[width=\textwidth]{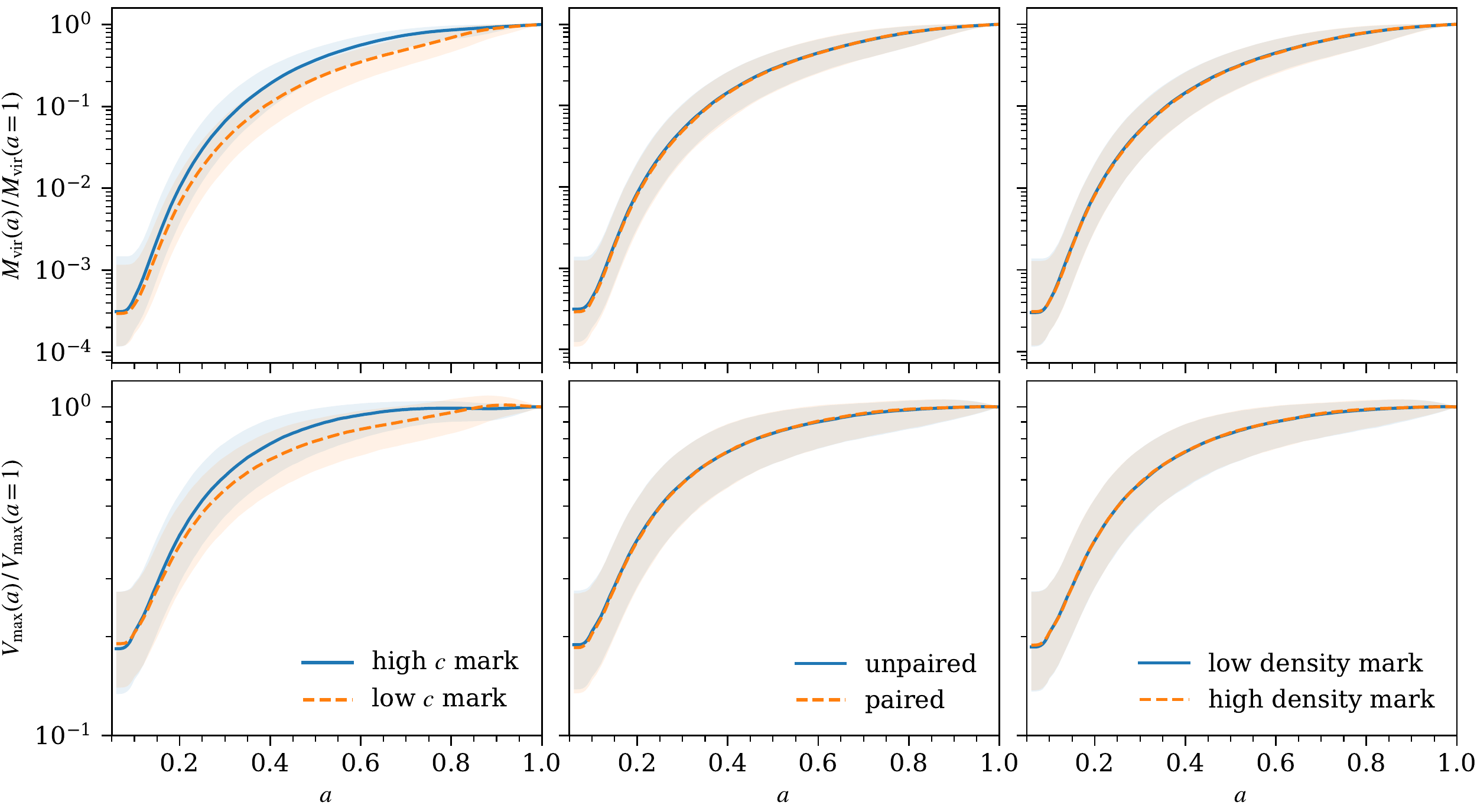}
\caption{\label{fig:history}%
Stacked main-branch assembly histories for halo mass $M(a)/M(a=1)$ (upper row) and for maximal circular velocity $\vmax(a)/\vmax(a=1)$ (lower row).
In each panel, the full sample of cluster-size haloes is split to show the difference between the stacked assembly histories of the two subsamples. 
In the three columns, the sample is split by concentration mark (left column; orange dashed line for the 50\% low-concentration haloes and blue solid line for the 50\% high-concentration),
by whether or not the haloes reside in pairs (middle column; orange dashed line for paired haloes and blue solid line for unpaired), 
and by large-scale density mark (left column; orange dashed line for the 50\% haloes in high-density regions and blue solid line for the 50\% haloes in low-density regions).
Here paired haloes are haloes that have neighbour haloes within 10\,\perh\,Mpc), and the large-scale density is calculated by summing the total mass$^\text{\ref{footnote:halo-proxy}}$ within a 20\,\perh\,Mpc-radius sphere.
For each subsample under consideration, the line shows the median mass assembly history, and the corresponding band shows the 16$^\text{th}$ and 84$^\text{th}$ percentiles.
This plot is made with the DarkSky-Gpc simulation. 
}
\end{figure*}

\subsection{{Full halo assembly histories and bias}}
\label{sec:history}

So far, we have inspected four different summary statistics of halo mass assembly history: \ahalf, \almm, \accrate, and \accratei, and \emph{none of them} exhibits any statistically significant secondary clustering bias. Given the similarity of halo mass assembly histories, which can be described well by one or two parameters, {one might \emph{speculate} that, for high-mass haloes, the entire halo mass assembly history is independent of the environment}. With the ``demography for paired haloes'' approach that we introduced in \autoref{sec:markdist}, we can directly inspect the mass assembly history for haloes {in different environments to further investigate the relationship between bias and assembly history.}

The upper row of \autoref{fig:history} shows the stacked (median) mass assembly histories for main-branch progenitors as a function of scale factor, $M(a)/M(a=1)$, using three different ways to split the full cluster-size halo sample into two subsamples. To produce \autoref{fig:history}, we used the DarkSky-Gpc simulation to obtain the full mass assembly histories for all distinct haloes that have a present-day mass $\mvir \geq 10^{14}\,\perh\,\msun$. On the left of \autoref{fig:history}, we split haloes by their concentration mark. In the middle, we split haloes by whether or not they are in pairs; here a ``paired halo'' is again defined as any distinct halo that has at least one other distinct halo closer than 10\,\perh\,Mpc. On the right, we split haloes by their {mass-normalized mark values of} large-scale matter density; we calculate the matter density by summing up the masses of all resolved haloes%
\footnote{We calculate the mass by summing up halo masses within 20\,\perh\,Mpc spheres because we do not have direct access to the full particle snapshot of DarkSky-Gpc. To make this approximation as close as the actual matter distribution, we use all distinct haloes identified in the DarkSky-Gpc halo catalogue, down to a minimal halo mass of $4.88 \times 10^{10}$\,\perh\,\msun{} (20 particles). These haloes, in total, contain 42.5\% of the total matter mass in the simulation box. For our purpose of ranking the large-scale matter densities, this method provides good approximation, as we have verified using independent simulations. \label{footnote:halo-proxy}}
within a 20\,\perh\,Mpc sphere around each distinct halo in our sample, and then calculate the mass-normalized mark values using the procedure outlined in \autoref{sec:marks}.

We can immediately see that the stacked mass assembly histories for paired and unpaired haloes are \emph{essentially identical}. The 16$^\text{th}$ and 84$^\text{th}$ percentiles also match between the two samples, indicating that the variety of possible assembly histories is quite similar for both paired and unpaired haloes. 
Furthermore, when the halo sample is split by the large-scale matter density around the haloes, haloes in denser regions have nearly identical stacked assembly history as those haloes in less dense regions, as shown in the upper right-hand panel of \autoref{fig:history}.
The lack of difference between the assembly histories of paired and unpaired haloes (or of haloes in high- and low-density regions) is \emph{not} caused by the stacking procedure. As the upper left-hand panel of \autoref{fig:history} shows, when the haloes are split by the concentration mark, there is a clear difference in the stacked assembly histories. The trend is consistent with our expectation, that high-concentration haloes form early. {One should also note that these two groups of haloes \emph{do} have different clustering biases, as we know the concentration bias does exist at this mass scale.}

We have also verified that the lack of difference in the assembly histories for haloes in different environments is insensitive to the radii used in the definitions of paired haloes and large-scale density (the result holds when using {5, 10, 20, 30, and 40}\,\perh\,Mpc), and is also insensitive to how we split the density mark (the result holds when selecting the 25\% or 10\% most extreme mark values).
In addition, we further inspect the history of the maximal circular velocity (\vmax) for main-branch progenitors, which represents the mass assembly history for the core of a halo. As the lower row of \autoref{fig:history} shows, we again find that the stacked \vmax{} histories of the paired and unpaired haloes, or of haloes in high- and low-density regions, are nearly identical. The concentration-split histories show a difference that is consistent with the difference in mass assembly history. We also find that the probability distributions of the number of major mergers (mass ratio between $1/3$ and 1) that happened along the main branch are also essentially the same for paired and unpaired haloes.

These findings are in good agreement with our main results and also with \citet{Gao2005} and \citet{2008MNRAS.389.1419L}. {At this mass scale ($\gtrsim 10^{14}$\,\perh\msun), haloes that are in different environments do not have significantly different assembly histories.
However, this does not guarantee that a correlation of any form between large-scale halo clustering and halo assembly history does not exist. Our results demonstrate that any such correlation is not evident either from the summary statistics that have been explored here or in the stacked assembly histories. The reasons for this phenomenon are twofold. First, at this mass scale, the strength of the secondary bias is small, and hence even if the $c-\ahalf$ relation were not biased for haloes in denser environments, the difference in the stacked assembly histories of haloes in different environments would still be modest. Secondly, as we discussed in \autoref{sec:demo}, the $c-\ahalf$ relation is slighted biased for haloes in denser environments, and it cancels out any remaining difference in the assembly histories for haloes in different environments.}

\begin{figure}
\centering\includegraphics[width=\columnwidth]{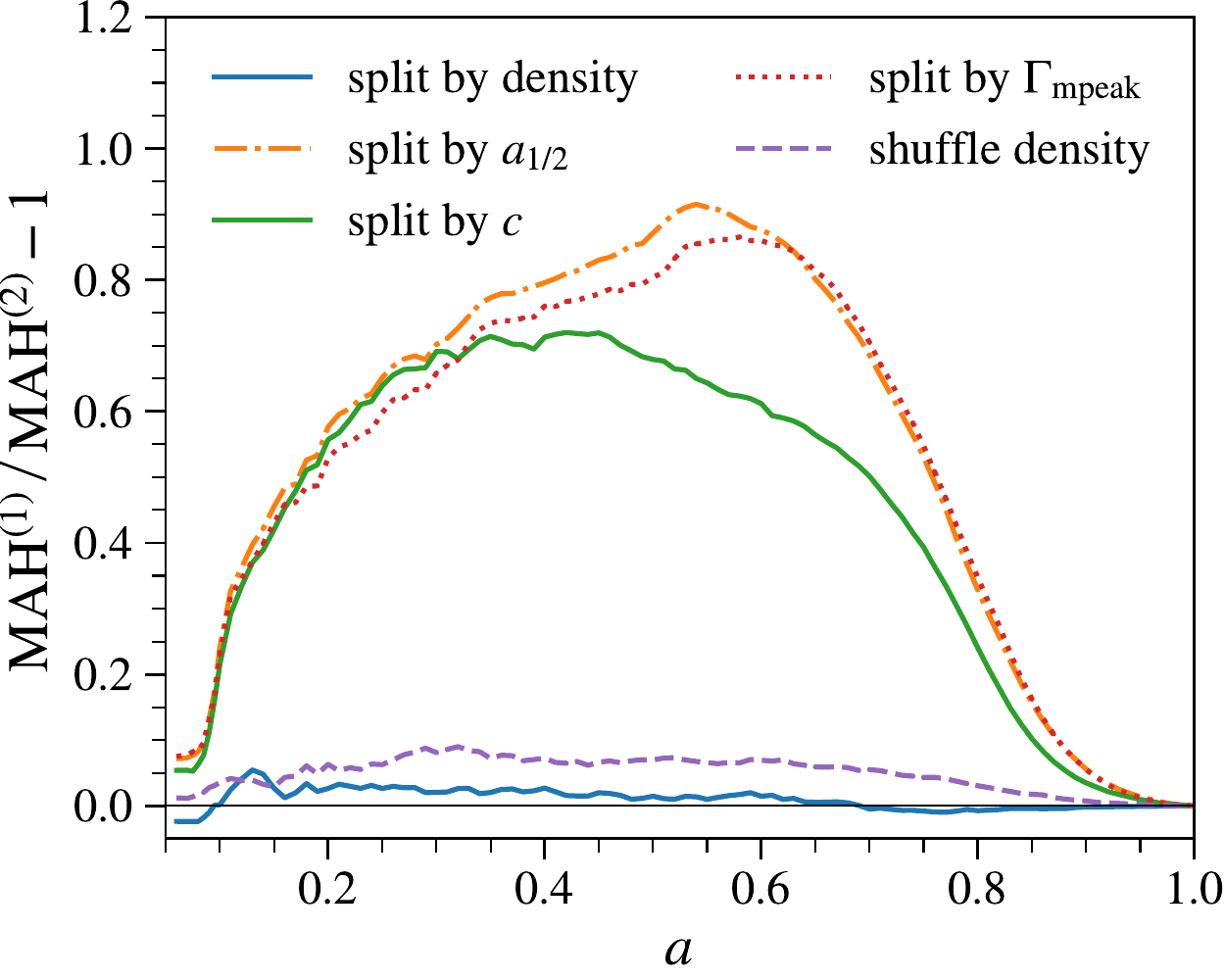}
\caption{\label{fig:history_diff}%
{Similar to \autoref{fig:history} but showing the relative difference in the stacked (median) main-branch mass assembly histories for two groups of haloes, for several different ways to split the haloes: by mark values of large-scale (20 \perh\,Mpc) density (blue solid, bottom-most), concentration (green solid, top-most), half-mass scale (orange dash--dotted), or accretion rate after peak mass (red dotted lines), and shuffled density (purple dashed line); all splits are made at the mark value of 0.5. The case when the haloes are split by shuffled density, each of the two groups has the same concentration distribution as in the case of split by large-scale density. 
This plot is made with the DarkSky-Gpc simulation. }
}
\end{figure}

{It is possible to estimate the relative sizes of these two effects. In \autoref{fig:history_diff}, we plot the relative differences between the stacked mass assembly histories of subgroups of haloes split by various halo properties. In each case, we split the samples into two equal-sized subgroups about a mark value of 0.5. For example, the green line in \autoref{fig:history_diff} shows the normalized difference between the blue and orange lines in the upper left-hand panel of \autoref{fig:history}. Splitting haloes by their 
\conc{}, \ahalf{}, or \accrate{} (green solid, orange dash--dotted, and red dotted lines respectively), yields significant differences in mass assembly histories, as one would expect. On the other hand, splitting haloes by large-scale density (blue solid line) yields assembly histories that are extraordinarily similar. This is another representation of our previous results. 
}

{
We can now explore what would be expected of the mass assembly difference between haloes selected by density due only to the fact that density is correlated with concentration. To compute this, we place haloes into narrow bins of concentration mark and then randomly shuffle the density marks. We then split haloes based upon the shuffled density mark. In this manner, the selection upon density becomes meaningless, but the two sub-populations have concentration mark distributions that are identical to the subgroups split on actual density. The mass accretion history difference constructed in this way is shown by the purple dashed line in \autoref{fig:history_diff}. This line shows the slightly different mass accretion histories of haloes due to the fact that selecting upon density also introduces small differences in the concentration distributions of the two halo subgroups. While this difference is small relative to selecting upon concentration or formation time directly, it is interesting that this difference is significantly larger than the difference in mass accretion histories induced by selecting on {\em actual} density. Despite the fact that selecting upon density also selects populations with slightly different concentration distributions, selecting upon {\em actual} density yields a nearly undetectable difference in mass accretion histories. The relation between concentration and mass accretion history is density dependent in such a way as to render the different mass accretion histories nearly identical. The difference in mass accretion histories exhibited by the purple dashed line in \autoref{fig:history_diff} is also what we would have observed if the secondary bias due to \ahalf{} or \accrate{} were as significant as the concentration bias.
}

{A common assumption has been that the secondary bias due to halo concentration is merely a consequence of the simple halo assembly bias. In other words, it has been commonly believed that secondary bias due to concentration was induced by the combination of assembly bias and the formation time--concentration relation. Given what we have learned here, the concentration bias seems more intriguing at this mass scale. In fact, one extremely interesting feature in \autoref{fig:history_diff} is that when the haloes are split by the concentration mark, the difference in the mass assembly history {is similar to} the difference when the haloes are split by half-mass scale or accretion rate {only at early time ($a < 0.35$)}. This suggests that, at this mass scale, halo concentration, as a summary statistic, {captures only early-time assembly history and is more correlated with large-scale bias.} We leave this intriguing phenomenon for future study. }

\subsection{Secondary biases due to subhalo properties}

In \autoref{fig:bias} and \ref{fig:bias-more} we see that the number of subhaloes and the average subhalo distance (both weighted and unweighted) give significant secondary bias, and that the peak maximal circular velocity of the largest subhalo, which at a fixed distinct halo mass represents the gap between the first subhalo and the distinct halo, also exhibits a modest secondary bias signal.
Similar to halo concentration, both the number of subhaloes and the average subhalo distance are correlated with the assembly history of the parent distinct halo, and both of them exhibit secondary bias despite the lack of assembly bias. This result is in good agreement with the findings in \citet{More2016}. 

The secondary halo biases due to these subhalo properties are particularly interesting because they are more likely to be directly observable; at this mass scale, most large subhaloes would host galaxies.
For example, if the way galaxies \emph{populate} subhaloes is not influenced by large-scale environment (although there is no clear evidence for such a dependence, it could exist), then the secondary bias due to subhalo occupation would directly translate to observable galaxy occupation bias (i.e., the dependence of central galaxy clustering on the member occupation or richness). Similarly, the average galaxy member distance will exhibit a bias signal if the galaxy--subhalo connection is not influenced by large-scale environment.

There are, however, complications to these potential observables. First, as we have discussed in \autoref{sec:galaxy-bias}, the galaxy occupation bias or the average galaxy distance bias does not directly translate to the halo concentration bias or the halo assembly bias (in fact, we have already shown the latter does not exist for cluster-size haloes).
Secondly, observationally, projection effects and redshift-space distortions can contaminate member assignment, and potentially produce artificial bias signal \citep{2016arXiv161100366Z,1702.01682}. 
Hence, one should be cautious when interpreting the galaxy occupation bias or the average galaxy distance bias.

\section{Conclusion}
\label{sec:conclusion}

In this study, we have revisited the complex phenomenon of secondary halo bias, which is commonly referred to as ``halo assembly bias,'' for cluster-size haloes ($\mvir \geq 10^{14}\,\perh\,\msun$) using large-volume simulations. As part of this investigation, we presented a novel approach to highlight secondary halo bias. Our approach was to study the demographics of paired and unpaired haloes, enabling us to determine whether or not the distributions of halo properties are different for haloes in pairs compared to the full halo sample. 

Using both halo two-point functions and paired halo demographics, we found that halo concentration, halo spin, number of subhaloes, and average subhalo distance all exhibit significant secondary halo biases at the cluster mass scale. Amongst these properties, halo spin exhibits the strongest secondary halo bias, in the sense that high-spin haloes cluster more strongly. Low-concentration haloes cluster more weakly than high-concentration haloes, a dependence that is the converse of concentration-dependent secondary halo bias at lower masses. Cluster-size haloes cluster more strongly as a function of both subhalo number and the average distance between subhaloes and the halo centre.

We have identified no statistically significant secondary halo bias at this mass scale for any of four halo properties that directly measure the mass assembly history: half-mass scale, accretion rate before peak mass, instantaneous accretion rate, and time of last major merger. We have found that the entire main-branch mass assembly histories of paired and unpaired haloes (or haloes in different large-scale densities) are statistically identical. {This suggests that the assembly histories of massive, cluster-size haloes do not correlate with their environments in a simple fashion. This is not equivalent to the statement that there exists no features of halo assembly histories that do correlate with environment and/or clustering strength. Indeed, some particular features of the halo assembly histories, such as those captured by halo concentration, may still correlate with the large-scale environment.}

With our halo demographic approach, we further investigated the seemingly contradictory result of the lack of secondary halo bias due to assembly history-related properties, given the clear correlation between halo concentration and halo assembly history. We have demonstrated that the correlation between two variables is, in general, not relevant to the question of whether or not the two variables will result in similar secondary halo clustering biases. If the conditional distributions of the two variables are altered, even only slightly, for paired haloes, then the two variables can easily result in very different secondary clustering biases. For instance, halo concentration and half-mass scale are similarly correlated for paired and unpaired haloes, yet paired haloes have a slightly lower concentration at a given half-mass scale, which results in their different secondary bias signals. Likewise, the fact that two variables (e.g., halo concentration and average subhalo distance) yield similar secondary clustering biases by no means implies  a correlation between the two variables. These statements have the following important consequence: Even when the presence of a galaxy in a halo is a function of a halo property that exhibits secondary halo bias, this does \emph{not} necessarily imply galaxy ``assembly bias.''

Our study has provided a comprehensive view of secondary halo biases for cluster-size haloes, and leads to a different perspective on secondary halo biases. In particular, we caution against use of the term ``assembly bias,'' particularly for cluster-size haloes.  The term ``assembly bias,'' when used to refer to secondary biases other than those due to the assembly history (e.g., concentration-dependent halo clustering), implies that all such biases have a common origin rooted in some aspect of the mass assembly histories of haloes. 
{While these different secondary biases may still all have some connections with the halo assembly history, those connections are more complex than simple correlations amongst different halo properties. 
In particular, the halo secondary bias due to concentration is not a direct consequence of the difference in bias between early- and late-forming haloes.}

Our results regarding halo assembly histories have an important consequence for the interpretation of halo clustering. At face value, our result is in qualitative agreement with early analytic studies of halo abundance and clustering using excursion set theory, which predicted no assembly bias (specifically, \citealt{Kaiser84,Cole1989,1991ApJ...379..440B,Mo1996,Sheth2001}; see \citealt{10.1142/S0218271807010511} for a review and subsequent developments). Yet, these predictions stem from ad hoc assumptions adopted for computational ease, rather than for any well-established physical reasons. Several authors have proposed more physically-motivated analytic interpretations of secondary halo bias based upon the assembly histories of haloes \citep{10.1142/S0218271807010511,Desjacques2007,Dalal2008}. {However, given our findings, these analytic interpretations of halo assembly bias do not manifest in all different kinds of secondary halo biases in a straightforward manner. Indeed, these interpretations contradict our finding that high-mass haloes cluster independently of halo formation time and other simple metrics of halo age. The explicit relations amongst large-scale environment, halo assembly history, and sundry internal halo properties, such as concentration and spin, remain unclear.}

While this study has demonstrated that a very complex phenomenology of secondary halo bias is mathematically possible, it has yet to provide a solid physical explanation for the existence of the concentration bias, the spin bias, the subhalo abundance bias, and the average subhalo distance bias. Previous attempts to explain these secondary biases that rely upon the existence of halo assembly bias in its restricted definition are not valid in this mass regime. It is also important to understand why the correlations between particular halo properties (e.g., between subhalo abundance and assembly history) depend on large-scale environment. We have not yet identified a plausible working theory that is able to explain all of the correlations. We hence leave this interesting problem for future study, with the hope that what we have laid out in this study will provide useful insights.

\section*{Acknowledgements}

The authors thank Andreas Berlind, Philipp Busch, Neal Dalal, \mbox{Andrew} Hearin, Ari Maller, Surhud More, Rita Tojeiro, \mbox{Li-Cheng} Tsai, Frank \mbox{van den Bosch}, Antonio \mbox{Villarreal}, and Kuan Wang for useful discussions. 
This research made use of the MDPL2 simulation; the authors gratefully acknowledge the Gauss Centre for Supercomputing e.V.\ (\http{www.gauss-centre.eu}) and the Partnership for Advanced Supercomputing in Europe (PRACE, \http{www.prace-ri.eu}) for funding the MultiDark simulation project by providing computing time on the GCS Supercomputer SuperMUC at Leibniz Supercomputing Centre (LRZ, \http{www.lrz.de}), and also thank Peter Behroozi for the direct access to the MDPL2 halo catalogues on SLAC servers. 
This research made use of the DarkSky-Gpc (\texttt{ds14\_b}) simulation, which was part of the Dark Sky Simulations (\http{darksky.slac.stanford.edu}) produced using an INCITE 2014 allocation on the Oak Ridge Leadership Computing Facility at Oak Ridge National Laboratory, a U.S.\ Department of Energy Office; the authors thank Sam Skillman, Mike Warren, Matt Turk, and other Dark Sky collaborators for their efforts in creating these simulations and for providing access to them. 
This research made use of computational resources at SLAC National Accelerator Laboratory, a U.S.\ Department of Energy Office; YYM and RHW thank the support of the SLAC computational team. 
YYM is supported by the Samuel P.\ Langley PITT PACC Postdoctoral Fellowship. 
ARZ is supported, in part, by grants AST 1516266 and AST 1517563 from the U.S.\ National Science Foundation as well as by the Pittsburgh Particle physics, Astrophysics, and Cosmology Center (PITT PACC) at the University of Pittsburgh. 
RHW received partial support from the U.S.\ Department of Energy contract to SLAC No.\ DE-AC02-76SF00515. 
This work was completed at the Kavli Institute for Theoretical Physics at the program ``The Galaxy--Halo Connection Across Cosmic Time,'' and supported in part by the National Science Foundation under Grant No.\ NSF PHY-1125915. 

This research made use of Python, along with many community-developed or maintained software packages, including
IPython \citep{ipython},
Jupyter (\http{jupyter.org}),
Matplotlib \citep{matplotlib},
NumPy \citep{numpy},
Pandas \citep{pandas},
and SciPy \citep{scipy}.
This research made use of NASA's Astrophysics Data System for bibliographic information.

\bibliographystyle{mnras}
\bibliography{references,software}

\appendix
\section{Correlations amongst all secondary halo properties}
\label{sec:appendix}

\begin{figure*}
\centering\includegraphics[width=\textwidth]{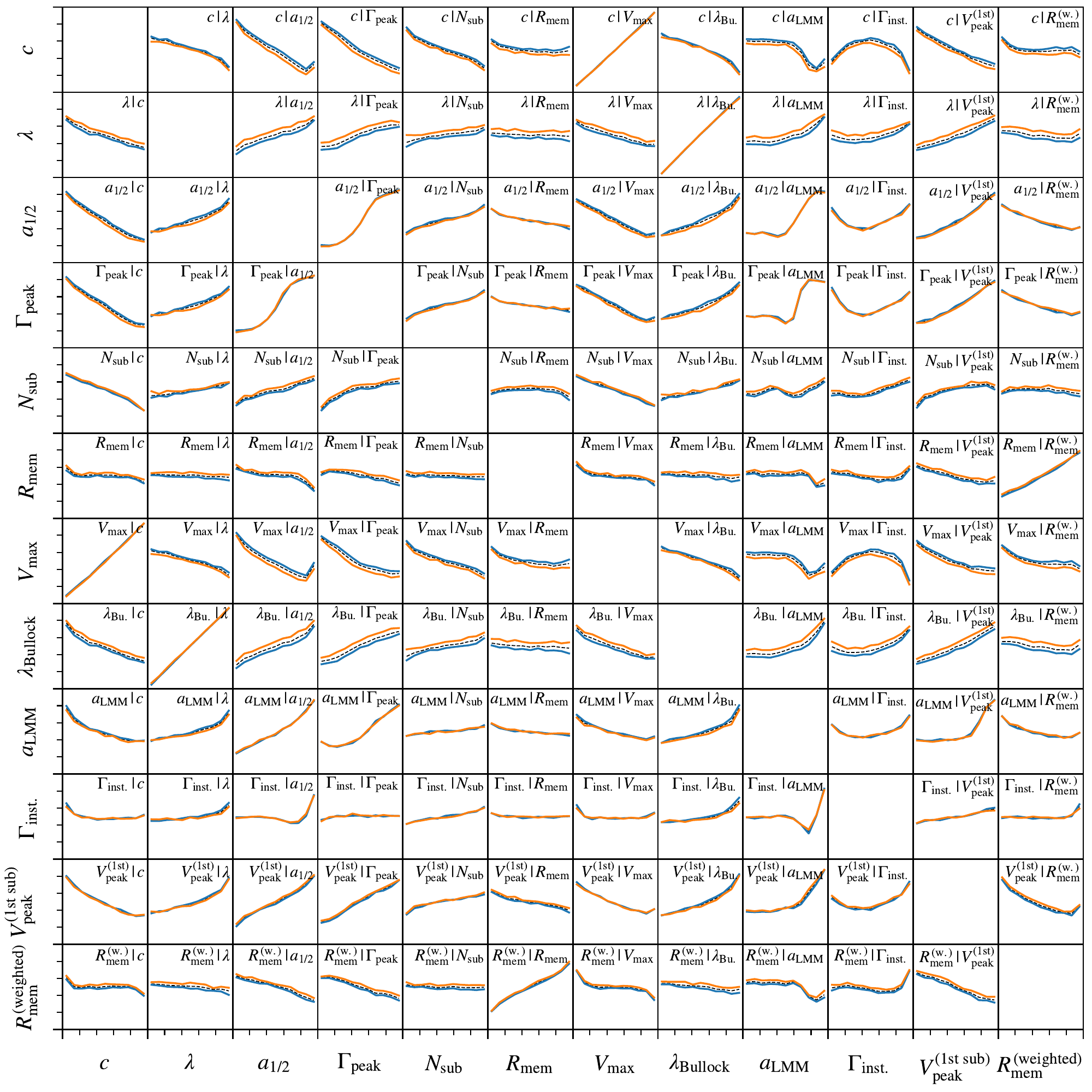}
\caption{\label{fig:conditional-mean}%
Similar to the right-hand panel of \autoref{fig:c-ahalf}, each cell of this matrix shows the mean value of the $y$-mark conditioned on the $x$-mark. For each cell, the $x$- and $y$-axes both go from 0 to 1 to show the mark value of the corresponding labels.
The black dashed, blue solid, and orange solid lines show the sample of all, paired, and unpaired haloes, respectively. 
The trend of the lines shows the correlation between the two properties, and the difference between the orange and blue lines indicates the difference in the conditional distributions for paired and unpaired haloes.}
\end{figure*}

For completeness, in \autoref{fig:conditional-mean} we show the correlations amongst the mark values of all 12 secondary halo properties considered for our cluster-size halo sample, by plotting the conditional mean as a summary statistic of the conditional distributions, and highlight the manner in which the conditional mean differs for paired and unpaired haloes (similar to the right-hand panel of \autoref{fig:c-ahalf}). 

Most of the results we have presented can also be observed in \autoref{fig:conditional-mean}. For example, we find that paired haloes have higher spin (\spin{}, second and eighth rows), when conditioned on any other halo property. This indicates both the observed secondary bias signal of \spin{} and the similar magnitude of \spin{}- and \spinb{}-biases.
We again see little split in the conditional mean between paired and unpaired haloes for all the assembly properties. We also see essentially no split for each pair of two halo properties that are similarly defined (e.g., concentration and maximal circular velocity).

\bsp	
\label{lastpage}
\end{document}